\documentclass[journal]{IEEEtran}

\ifCLASSINFOpdf
\else
\fi

\usepackage{amsthm,amsmath,amssymb}
\usepackage{tabularx}
\usepackage{booktabs}
\usepackage{graphicx}
\usepackage{url}
\usepackage{color}
\usepackage{ulem}

\theoremstyle{plain}% Theorem-like structures
\newtheorem{theorem}{Theorem}

\newtheorem{remark}[theorem]{Remark}

\newcommand{\R}{\mathbb{R}}
\newcommand{\stackcell}[2]{\begin{tabular}{@{}c@{}}#1\\#2\end{tabular}}

\newcommand{\bgamma}{\boldsymbol{\gamma}}
\newcommand{\balpha}{\boldsymbol{\alpha}}
\newcommand{\bX}{\boldsymbol{X}}

\newcommand{\bU}{\boldsymbol{U}}
\newcommand{\bV}{\boldsymbol{V}}
\newcommand{\bS}{\boldsymbol{S}}
\newcommand{\bs}{\boldsymbol{s}}

\begin{document}
\title{Exact Semiparametric Inference and Model Selection for Load-Sharing Systems}

\author{Fabian Mies and Stefan Bedbur% <-this % stops a space
\thanks{Both authors are with RWTH Aachen University, Institute of Statistics, Germany, D-52062 Aachen.}%
\thanks{Email: mies@stochastik.rwth-aachen.de, bedbur@isw.rwth-aachen.de}
\thanks{\textcopyright~2019 IEEE.  Personal use of this material is permitted.  Permission from IEEE must be obtained for all other uses, in any current or future media, including reprinting/republishing this material for advertising or promotional purposes, creating new collective works, for resale or redistribution to servers or lists, or reuse of any copyrighted component of this work in other works.}
\thanks{Digital Object Identifier 10.1109/TR.2019.2935869}}

\markboth{To appear in: IEEE Transactions on Reliability}%
{Mies, Bedbur: Exact Semiparametric Inference Based on SOSs}



\maketitle

\begin{abstract}
	As a specific proportional hazard rates model, sequential order statistics can be used to describe the lifetimes of load-sharing systems. Inference for these systems needs to account for small sample sizes, which are prevalent in reliability applications.
	By exploiting the probabilistic structure of sequential order statistics, we derive exact finite sample inference procedures to test for the load-sharing parameters and for the nonparametrically specified baseline distribution, treating the respective other part as a nuisance quantity. This improves upon previous approaches for the model, which either assume a fully parametric specification or rely on asymptotic results. Simulations show that the tests derived are able to detect deviations from the null hypothesis at small sample sizes. Critical values for a prominent case are tabulated.
\end{abstract}


% \fi
\IEEEpeerreviewmaketitle

\section{Introduction}
\IEEEPARstart{I}{n} many technical systems, redundant components are added to maintain operability in case of component failures. A $k$-out-of-$n$ system, for instance, consists of $n$ identical components and remains active as long as $k$ components are operating. The failure times of the components are typically described by independent and identically distributed (iid), non-negative random variables distributed according to some baseline cumulative distribution function (cdf) $F$, in the case of which the $(n-k+1)$-th order statistic based on $F$ represents the system's lifetime. In practice, however, the assumption of independence is often inadequate. If, for example, all active components equally share the total load of the system, the failure of any single component is likely to increase the stress put on the remaining ones. To model the lifetimes of such systems and their components, sequential order statistics (SOSs) have been introduced in \cite{kamps1995concept,Kam1995b}, where upon failure of some component the underlying lifetimes distribution may change. In a setup with proportional hazard rates, SOSs are introduced via some baseline cdf $F$ and positive parameters $\alpha_1,\dots,\alpha_n$, say, which describe the change of the underlying lifetime distributions due to load-sharing effects, for instance. Hence, $\alpha_1,\dots,\alpha_n$, which are referred to as load-sharing parameters in what follows, may be interpreted as an attribute of the interoperation with the technical system, whereas $F$ describes the quality of the individual components. For an extensive account on the model of SOSs including structural properties and inference, we refer to \cite{CraKam2001b}. 

Based on observations of (multiple) samples of SOSs, inference has been performed for the load-sharing parameters $\alpha_1,\dots,\alpha_n$ and/or for the baseline cdf $F$. By assuming $F$ to be known or in some parametric family, inferential results for load-sharing and distribution parameters of SOSs are provided by, e.g., \cite{BalBeuKam2008,BalBeuKam2011,BedBeuKam2012,BedBurKam2016,BedLenKam2013,BeuKam2009,cramer1996sequential,CraKam2001a,CraKam2001b}. On the other hand, the literature on nonparametric inferential methods for SOSs is rather scarce. In the more general multiplicative intensity model, which comprises the SOSs model, the Cox regression model (see \cite{cox1972regression}), and various types of censoring, 
\cite{aalen1978nonparametric} considers an approach based on counting processes and suggests a nonparametric estimator for the unknown intensity function. This estimator is discussed and contrasted in different aspects by \cite{jacobsen1984maximum}. Statistical procedures in the more specific proportional hazard rate setting of SOSs take the baseline cdf $F$ and the load-sharing parameters $\alpha_1,\dots,\alpha_n$ as unknown quantities. In this semiparametric framework, \cite{kvam2005estimating} study the asymptotic behaviour of the Nelson-Aalen estimator in combination with a profile-likelihood approach for the load-sharing parameters. The results have been applied in
\cite{beutner2008nonparametric,Beu2010b,beutner2010nonparametric} to construct asymptotic test statistics for the nonparametric and the parametric part of the SOSs model.

The restriction of the general multiplicative intensity model to SOSs based on $F$ and $\alpha_1,\dots,\alpha_n$ allows for exploitation of additional structural properties. Here, our key observation is that the estimators studied by \cite{kvam2005estimating} and \cite{beutner2008nonparametric} give rise to several statistics which are distribution-free with respect to (w.r.t.) $F$. These statistics can be used to perform exact statistical inference based on finite samples which is, in particular, relevant for applications in life-testing, where experimental data sets are typically small. Based on the key result, we design exact statistical tests for the load-sharing parameters $\alpha_1,\dots,\alpha_n$ as well as for the baseline cdf $F$ with the respective other part considered an (unknown) nuisance parameter. The critical values of these tests do not admit an analytical expression, but can be approximated to arbitrary precision by Monte Carlo methods.

The remainder of this paper is structured as follows. In Section \ref{sec:model-spec}, we review the model of SOSs based on $F$ and $\alpha_1,\dots,\alpha_n$ along with some basic results. In Section \ref{sec:model-check}, it is shown that the estimators in 
\cite{kvam2005estimating} and \cite{beutner2008nonparametric} are distribution-free w.r.t. the baseline cdf $F$.
This finding is then utilized to derive exact statistical tests within the semiparametric SOSs model. In Section \ref{ss:mc}, tests for the load-sharing parameters are proposed, where $F$ is assumed to be unknown; in particular, these tests allow to check for static intensities and may be considered tests for model-selection in the sense whether order statistics may serve as a simpler model for the data. A simulation study is carried out to compare the tests in terms of power, and tables of critical values are addressed. Goodness-of-fit tests for the baseline cdf $F$ are provided in Section \ref{ss:gof}, where the load-sharing parameters are supposed to be known or unknown. In the latter case, the performance of the tests is studied based on another power study. Section \ref{s:asym} deals with size and power comparisons to previous asymptotic tests from the literature, and the exact tests are applied to a real data example in Section \ref{s:realdata}. Finally, Section \ref{sec:concl} gives the conclusion being followed by a section with deferred proofs.

\section{Model and Basic Properties}\label{sec:model-spec}

First, we illustrate the model by means of an example. We consider an $(n-r+1)$-out-of-$n$ system, the component lifetimes of which are described by SOSs $X^{(1)}\leq\dots\leq X^{(r)}$ based on some absolutely continuous cdf $F$ with hazard rate $\lambda_F(t)=-d/dt\ln(1-F(t))$, $t>0$, and positive parameters $\alpha_1,\dots,\alpha_n$. Then, all components start operating at hazard rate $\alpha_1\lambda_F$ until the first failure occurs. Once the $k$-th component fails at time $X^{(k)}$, $1\leq k\leq r-1$, the hazard rate of the remaining $n-k$ components changes from $\alpha_k\lambda_F$ to $\alpha_{k+1}\lambda_F$. Finally, the $r$-th component failure time $X^{(r)}$ is also the system's lifetime after which no further failure times are recorded. In that sense, we have type-II right censored observations $X^{(1)}\leq\dots\leq X^{(r)}$ from a sample of size $n$.

There are various equivalent definitions of SOSs in the literature making this intuitive construction mathematically precise (see \cite{kamps1995concept,Kam1995b,CraKam2003}). Here, we employ a formulation of the model in terms of counting processes, analogous to that in \cite{kvam2005estimating} and \cite{beutner2008nonparametric}; in particular, we assume that $\alpha_1=1$ to guarantee identifiability of the underlying parameters in a semiparametric setting.

Let $N=(N(t))_{t\geq0}$ be a counting process with jump times $X^{(1)} \leq \ldots \leq X^{(r)}$, i.e.,
\begin{align*}
	N(t)\,=\,\sum_{j=1}^r \mathbf{1}_{\{X^{(j)}\leq t\}}\,,\quad t\geq 0\,,
\end{align*}
where $\mathbf{1}_A$ denotes the indicator function of a set $A$. According to \cite{jacod1975multivariate}, $N$ is uniquely determined by its (cumulative) stochastic intensity process $\Lambda=(\Lambda(t))_{t\geq0}$, which is an increasing process, predictable w.r.t.\ the natural filtration $\sigma(N(s), s\leq t)$, $t\geq0$, such that $N(t)-\Lambda(t)$, $t\geq0$, is a martingale. Then, increasingly ordered random variables $X^{(1)},\dots,X^{(r)}$ are distributed as the first $r\,(\leq n)$ SOSs based on some absolutely continuous cdf $F$ and positive parameters $\alpha_1,\dots,\alpha_n$ if the cumulative intensity process of the counting process $N$ is given by
\begin{align*}
	\Lambda(t) \,=\,  
	\int_0^t \lambda_F(s)\,\gamma_{N(s-)+1}\, ds\,,\quad t\geq0\,,
\end{align*} 
with $\gamma_{k}=(n-k+1)\alpha_k$, $1\leq k\leq r$, and $\gamma_{r+1}=0$, where we use the notation $N(s-)=\lim_{u\uparrow s}N(u)$. The particular case $r=n$ corresponds to a complete sample of SOSs based on $F$ and $\alpha_1,\dots,\alpha_n$. In the definition of the $\gamma$'s, the factors account for the fact that, for $1\leq k\leq r$, $X^{(k)}$ is the first failure time of the remaining $n-k+1$ identical components still at work.
Throughout the article, we use the description via the $\gamma$'s and $\alpha$'s interchangeably, since the former simplifies expressions while the latter is easier to interpret. Moreover, let $\balpha=(\alpha_1,\dots,\alpha_r)$ and $\bgamma=(\gamma_1,\dots,\gamma_r)$ for brevity. Finally, note that for $\balpha=(1,\dots,1)$ and $\bgamma=(n,n-1,\dots,n-r+1)$, respectively, $X^{(1)},\dots,X^{(r)}$ are distributed as the first $r\,(\leq n)$ ordinary order statistics based on $F$. 

As mentioned in the introduction, we are interested in statistical inference for the load-sharing parameter $\balpha$ and the baseline cdf $F$. If $F$ is known and absolutely continuous,  the joint distribution of $X^{(1)},\dots,X^{(r)}$ forms a regular exponential family in $\balpha\in(0,\infty)^r$ such that useful inferential results are near at hand (see \cite{BedBeuKam2012}). Moreover, in that case, inference for $\balpha$ may be based on the fact that the transformed random variables $F(X^{(j)})$, $j=1,\ldots, r$, are distributed as the first $r\,(\leq n)$ SOSs based on the standard uniform cdf $U(x)=x, x\in [0,1]$, and parameters $\alpha_1,\dots,\alpha_n$ (see, e.g., \cite{kamps1995concept,Kam1995b}). As a particular finding, the maximum likelihood estimator (MLE) $\hat{\balpha}^\text{ML}$ of $\balpha$ based on $M$ iid samples $(\textbf{X}_i^{(1)},\ldots, \textbf{X}_i^{(r)})$, $1\leq i\leq M$, uniquely exists and has components
\begin{equation}
	\hat\alpha_j^\text{ML}\,=\,\frac{M}{\sum_{i=1}^M S_i^{(j)}}\,,\quad 1\leq j\leq r\,,\label{eqn:alphamle}
\end{equation}
with statistics
\begin{align*}
S_i^{(j)}&\,=\,-(n-j+1)\ln\left(\frac{1-F(X^{(j)}_i)}{1-F(X^{(j-1)}_i)}\right),
\end{align*}
for $1\leq j\leq r$ and $1\leq i\leq M$. To simplify notation, we set $F(X_i^{(0)})\equiv0$, $1\leq i\leq M$.
It is known that the statistics $\alpha_j S_i^{(j)}$, $1\leq j\leq r$, $1\leq i\leq M$, are iid having a standard exponential distribution (see, e.g., \cite{cramer1996sequential}). In particular, $\hat{\balpha}^\text{ML}$ is distribution-free w.r.t. the baseline cdf $F$, though the evaluation of the estimator requires knowledge of $F$.

\section{Statistical Tests and Model-Selection}\label{sec:model-check}

We are interested in statistical inference for the load-sharing parameter $\balpha$ and the baseline cdf $F$ in the semiparametric setting of SOSs.
For this, let $F$ be absolutely continuous with $F(0)=0$. The class of cdfs $F$ satisfying this assumption will be denoted by $\mathcal{F}$. Yet, an estimator for $F\in\mathcal{F}$ may be as well a discrete cdf, as suggested by \cite{kvam2005estimating}. Therein, the nonparametric Nelson-Aalen estimator for $F$ is studied, which is based on counting processes 
\begin{align*}
	N_i(t)\,=\,\sum_{j=1}^r \mathbf{1}_{\{X_i^{(j)}\leq t\}}\,,\quad t\geq 0\,,\quad 1\leq i\leq M\,,
\end{align*}
corresponding to $M$ iid samples $\bX_i=(X_i^{(1)},\dots,X_i^{(r)})$, $i=1,\ldots, M$, where $X_i^{(j)}$ describes the $j$-th failure time in system $i$, $1\leq i\leq M$, $1\leq j\leq r$. For known $\bgamma$ (or, equivalently, for known load-sharing parameter $\balpha$), the derived estimator $\hat{F}_{\bgamma}$ is a step function given by 
\begin{align}
	\label{eqn:F-kvam}\hat{F}_{\bgamma}(t)\,&=\,1-\prod_{X_i^{(j)}\leq t} \left(1-Y_i^{(j)}\right)\,,\quad t\geq0\,,\\[1ex]
	\label{eqn:q}\text{with}\qquad Y_i^{(j)}\,&=\,\left( \sum_{k=1}^M \gamma_{N_k(X_i^{(j)}-)+1} \right)^{-1},
\end{align}
for $1\leq j\leq r$ and $1\leq i\leq M$.

If $\bgamma$ is unknown, \cite{kvam2005estimating} suggest to, first, compute the estimator 
\begin{equation}
	\label{eqn:gammaprofile}\hat{\bgamma}^\text{PL}\,=\,\arg\max_{\bgamma} L(\bgamma, \mathbf{X})
\end{equation}
for $\bgamma$, being obtained by maximizing the profile likelihood 
\begin{equation}\label{eq:profile}
	L(\bgamma, \mathbf{X})\,=\,\prod_{i=1}^M \prod_{j=1}^r \frac{\gamma_{j}}{\sum_{k=1}^M \gamma_{N_k(X_i^{(j)}-)+1}}
\end{equation}
w.r.t. $\bgamma$, where $\mathbf{X}=(X_i^{(j)})_{1\leq i\leq M,1\leq j\leq r}\in(0,\infty)^{M\times r}$ denotes the data matrix. Then, $\hat{\bgamma}^{PL}$ is inserted for $\bgamma$ in formula (\ref{eqn:q}), which gives an estimator  $\hat{F}_{\hat{\bgamma}^\text{PL}}$ of $F$. Here, the maximum in formula (\ref{eqn:gammaprofile}) is taken w.r.t. $\bgamma\in(0,\infty)^r$ satisfying $\gamma_1=n$, since $\alpha_1=1$ by assumption. That is, we have $\hat{\gamma}_1^{\text{PL}}=n$ in the following.

Similar likelihood based estimators exist for more general recurrent event models (see \cite{PenHol2004,PenSlaGon2007,Pen2016}). Here, the consistency of the estimators depends on the exact specification of the model (see \cite{BeuBorDoy2017}). For the SOSs model studied in this paper, the presented estimators are consistent as the sample size $M$ tends to infinity, and corresponding central limit theorems are available (see \cite{kvam2005estimating}). The limiting Gaussian distributions can be used to design an asymptotic test for the baseline cdf or the load-sharing parameter (see \cite{beutner2008nonparametric,beutner2010nonparametric}). However, sample sizes in life-testing applications are often small, in the case of which asymptotic tests might not be appropriate. For example, if $X^{(r)}_i > X^{(r-1)}_j$ for all $i,j=1,\ldots, M$, it can be seen that the profile likelihood estimator degenerates, i.e., $\hat{\gamma}^{PL}_r = 0$. This situation occurs with probability tending to zero as $M\to\infty$ and is thus not reflected in the asymptotic results. In the following, we propose test procedures which admit exact significance levels for finite sample sizes.

\subsection{Testing the load-sharing parameter}\label{ss:mc}

The profile likelihood in formula (\ref{eq:profile}) depends on the data $\mathbf{X}$ only via
\begin{equation*}
N_k(X_i^{(j)}-)\,=\,\sum_{l=1}^r \mathbf{1}_{\{X_k^{(l)} < X_i^{(j)}\}}\,,\quad 1\leq k\leq M\,.
\end{equation*}
In particular, it only depends on the ordering of all observed failure times in $\mathbf{X}$ and not on the values themselves. This is reminiscent of the partial likelihood in the Cox regression model; see \cite{cox1972regression}. Since any (absolutely) continuous cdf $F$ is strictly increasing on its support $\{x\in\R:F(x+\delta)>F(x-\delta)\,\forall\delta>0\}$, it holds almost surely (a.s.)
\begin{equation}\label{eq:F}
	X_i^{(j)} < X_k^{(l)} \quad\iff\quad 
	U_i^{(j)} < U_k^{(l)}\,,
\end{equation}
where we define the random variables $U_i^{(j)}$ by
\begin{equation}\label{eq:Y}
U_i^{(j)}\,=\,F(X_i^{(j)})\,,\quad 1\leq j\leq r\,,\quad 1\leq i\leq M\,.
\end{equation}
Hence, we have $L(\bgamma, \mathbf{X})=L(\bgamma, \mathbf{U})$ a.s. and $\hat{\bgamma}^\text{PL}=
\arg\max_{\bgamma} L(\bgamma, \mathbf{U})$ a.s. with $\mathbf{U}=(U_i^{(j)})_{1\leq i\leq M,1\leq j\leq r}$, where
$\bU_i=(U_i^{(1)},\dots,U_i^{(r)})$, $1\leq i\leq M$, are iid random vectors distributed as the first $r\,(\leq n)$ SOSs based on the standard uniform cdf and parameters $\alpha_1,\dots,\alpha_n$, each. The representation $X_1^{(k)}=F^{-1}(U_1^{(k)})$, $1\leq k\leq n$, with quantile function $F^{-1}$ corresponding to $F$ and SOSs $U_1^{(1)},\dots,U_1^{(n)}$ based on the standard uniform cdf is known to hold true, even for continuous $F$ (see \cite{KamCra2001,CraKam2003}). This and related properties will turn out to be useful throughout the manuscript.

We thus obtain the following property, which is analogous to the properties of the regression estimator in the Cox model; see \cite{cox1972regression}.

\begin{theorem}
	The estimator $\hat{\bgamma}^\text{PL}$ is distribution-free w.r.t.\ the baseline cdf $F\in\mathcal{F}$.
\end{theorem}

Moreover, $\hat{\bgamma}^\text{PL}$ has finite support, i.e., it attains only finitely many different values with positive probability, one for each admissible ranking of the quantities $X_i^{(j)}$ in $\mathbf{X}$ (see formulas (\ref{eqn:gammaprofile}) and (\ref{eq:profile})). By specifying $\bgamma$, the distribution of $\hat{\bgamma}^\text{PL}$ can be obtained by means of Monte Carlo simulations. This finding can be used, e.g., to construct a test statistic $T_{\bgamma^{(0)}}$ based on $\hat{\bgamma}^\text{PL}$ for the simple null hypothesis $H_0:\bgamma=\bgamma^{(0)}$ subject to a desired exact significance level, where $\bgamma^{(0)}=(n,\gamma_2^{(0)},\dots,\gamma_r^{(0)})\in(0,\infty)^r$ is a fixed parameter vector. 
For instance, one could employ the log-likelihood-ratio statistic 
\begin{align}\label{eq:lrt}
\text{LR}_{\bgamma^{(0)}}\,=\,\ln L(\hat{\bgamma}^\text{PL}, \mathbf{X})-\ln L(\bgamma^{(0)}, \mathbf{X})
\end{align}
or the statistic 
\begin{align}\label{eq:Wald}
\text{W}_{\bgamma^{(0)}}\,=\,\sum_{j=1}^r \left(\frac{\hat{\gamma}_j^{\text{PL}}}{\gamma_j^{(0)}}-1\right)^2\,,
\end{align}
which, when replacing $\hat{\bgamma}^{\text{PL}}$ by $\hat{\bgamma}^{\text{ML}}$, coincides with the Wald statistic in case of a known baseline cdf $F$ (see \cite{BedBeuKam2014}). Here, both tests reject the null hypothesis for large values of the corresponding test statistic.

\begin{figure*}[ht!]
\centering
	\begin{minipage}{0.4\textwidth}
	\center{$\balpha=(1,1.4,1.8,2.2)$}
		\includegraphics[width=0.95\textwidth]{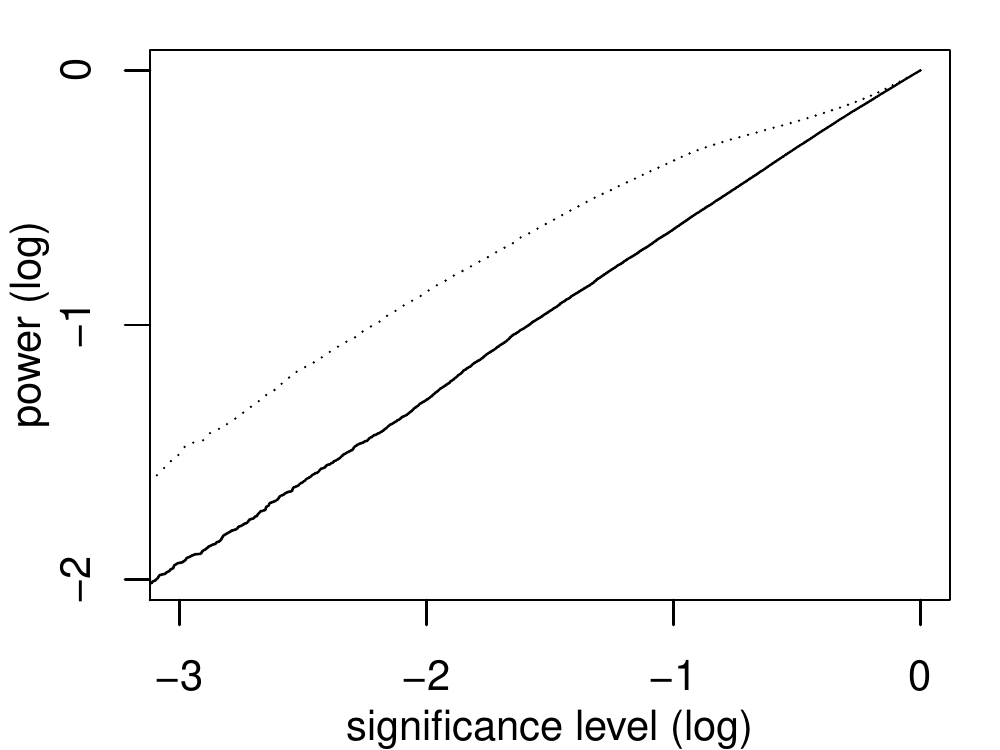}
	\end{minipage}
	\begin{minipage}{0.4\textwidth}
			\center{$\balpha=(1,1.25,1.9,4.1)$}
		\includegraphics[width=0.95\textwidth]{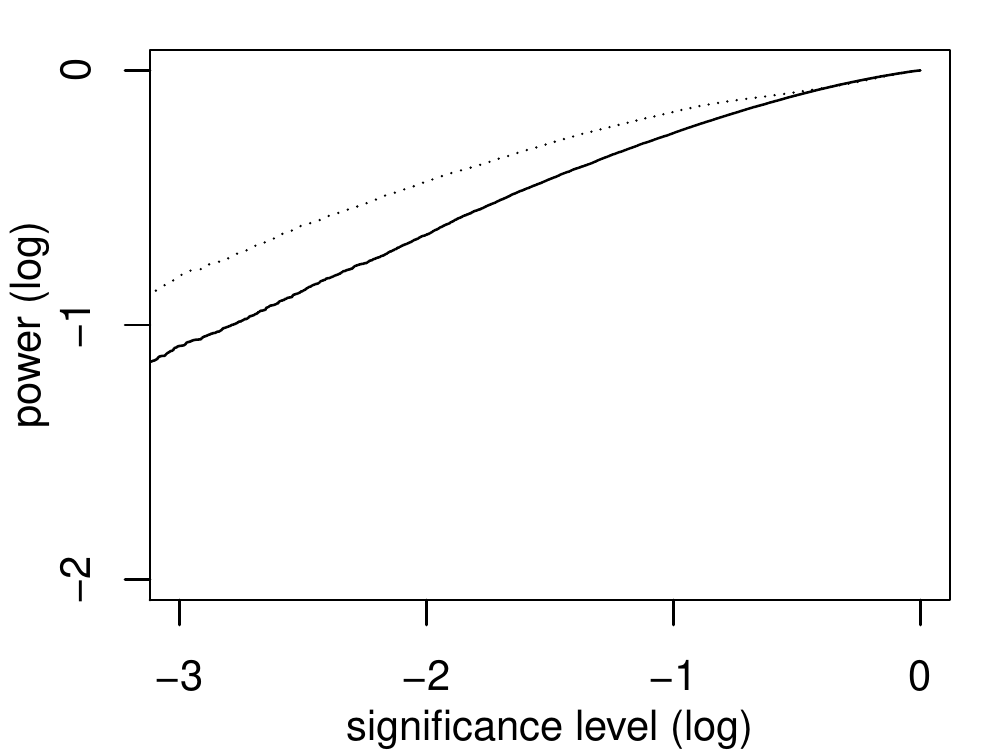}
	\end{minipage}
	
	\vspace{5mm}
	\begin{minipage}{0.4\textwidth}
			\center{$\balpha=(1,1,1,4)$}
%		\vspace{-3mm}
		\includegraphics[width=0.95\textwidth]{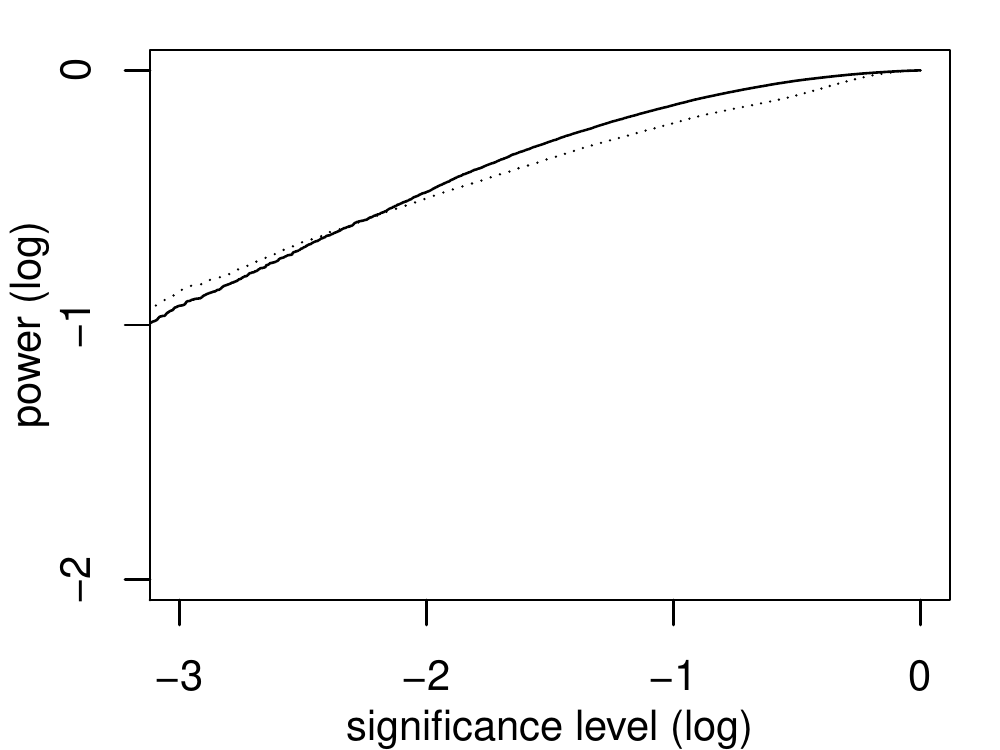}
	\end{minipage}
	\begin{minipage}{0.4\textwidth}
			\center{$\balpha=(1,1,3.3)$}
%		\vspace{-3mm}
		\includegraphics[width=0.95\textwidth]{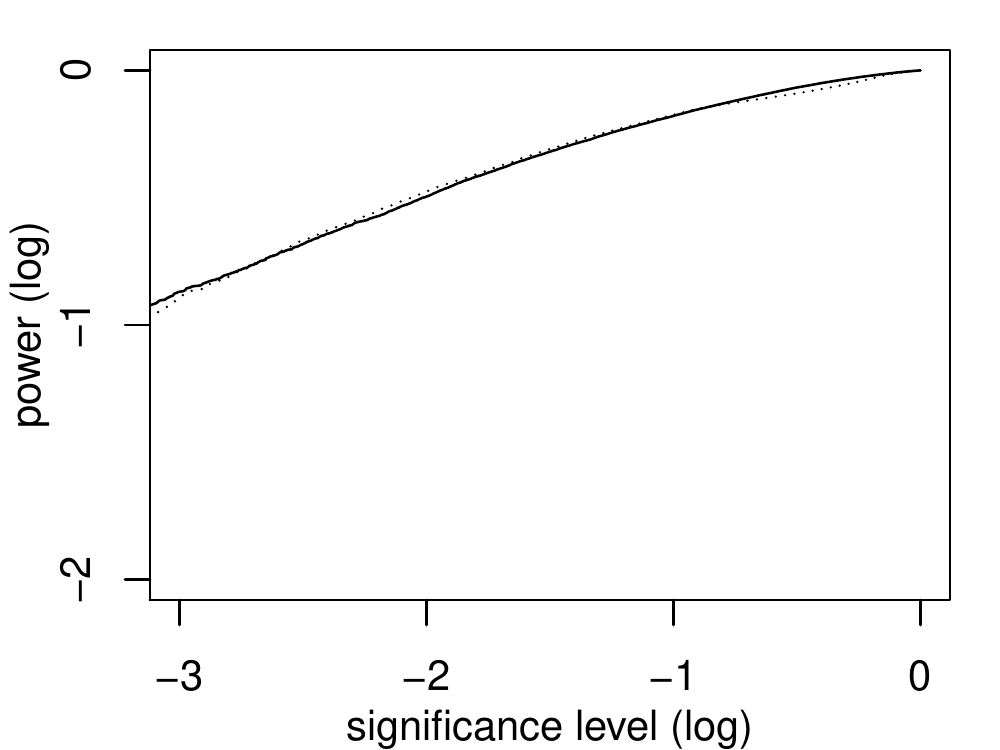}
	\end{minipage}
	\caption{Power at different alternatives $\balpha$ of the tests based on $\text{LR}_{\bgamma^{(0)}}$ (solid line) and $\text{W}_{\bgamma^{(0)}}$ (dotted line) for $r=n=4$, $M=10$, and $\balpha^{(0)}=(1,1,1,1)$ respectively $\bgamma^{(0)}=(4,3,2,1)$ as a function of the significance level (log-log plot based on $10^5$ realizations of every test statistic for every significance level).}
\label{fig:power1}
\end{figure*}

Of special interest is to check for static intensities, which can be done by setting $\bgamma^{(0)}=(n,\dots,n-r+1)$ with corresponding load sharing parameter $\balpha^{(0)}=(1,\dots,1)$. This test allows for model selection in the sense whether, based on the data, the model of ordinary order statistics has to be rejected in favour of the more flexible SOSs model, which may take load-sharing effects into account. As a function of the significance level, Figure \ref{fig:power1} shows the power of the above tests for $r=n=4$ and $M=10$ at different alternatives, which roughly correspond to the cases of either a linear or exponential increase of stress, or to periods of constant stress. For a better interpretation, the alternatives are presented by using the parametrization via the $\alpha$'s. The depicted results indicate a similar power of the tests, where the likelihood ratio test seems to be slightly inferior at the alternatives with pairwise distinct $\alpha$'s. For both tests, simulated critical values are listed in Tables \ref{tab:crits-LR} and \ref{tab:crits-WP} for different values of $r$, $n$, and $M$ subject to a significance level of 10\% and 5\%, respectively. All reported critical values are based on $10^5$ simulations.

\begin{table*}[tb]
\centering
	\begin{tabularx}{\textwidth}{r|*{8}{X}|*{8}{X}}
	 \toprule & \multicolumn{8}{c}{$\mathbf{M=5}$} &\multicolumn{8}{|c}{$\mathbf{M=10}$}\\ \midrule
	$\mathbf{n}$ & 3 & 4 & 5 & 6 & 7 & 8 & 9 & 10 & 3 & 4 & 5 & 6 & 7 & 8 & 9 & 10 \\
	  \midrule
	  $\mathbf{r}$ & & & & & & & & & & & & & & & & \\
	3 & \stackcell{ 2.80}{ 3.30} & \stackcell{1.81}{2.52} & \stackcell{1.52}{2.22} & \stackcell{1.27}{1.99} & \stackcell{ 1.10}{ 1.80} & \stackcell{.911}{1.64} & \stackcell{.769}{1.45} & \stackcell{.680}{1.39} & \stackcell{ 2.50}{3.27} & \stackcell{1.87}{2.58} & \stackcell{1.63}{2.33} & \stackcell{1.36}{2.07} & \stackcell{ 1.20}{1.91} & \stackcell{1.05}{1.75} & \stackcell{.901}{ 1.60} & \stackcell{.808}{1.52} \\[7pt]  
	  4 &   & \stackcell{3.69}{4.36} & \stackcell{2.48}{3.29} & \stackcell{2.12}{2.92} & \stackcell{1.78}{ 2.60} & \stackcell{1.45}{2.27} & \stackcell{1.23}{2.08} & \stackcell{1.02}{1.85} &   & \stackcell{3.36}{4.19} & \stackcell{2.59}{ 3.40} & \stackcell{2.25}{3.04} & \stackcell{1.93}{2.73} & \stackcell{1.68}{2.48} & \stackcell{1.43}{2.23} & \stackcell{1.27}{2.09} \\[7pt]  
	  5 &   &   & \stackcell{4.49}{5.29} & \stackcell{3.11}{4.02} & \stackcell{2.68}{3.58} & \stackcell{2.29}{3.19} & \stackcell{1.96}{2.85} & \stackcell{1.59}{2.53} &   &   & \stackcell{4.18}{5.07} & \stackcell{3.24}{4.14} & \stackcell{2.86}{3.76} & \stackcell{2.56}{3.44} & \stackcell{2.22}{3.11} & \stackcell{1.87}{2.78} \\[7pt]  
	  6 &   &   &   & \stackcell{5.28}{6.17} & \stackcell{3.72}{ 4.70} & \stackcell{3.27}{4.25} & \stackcell{2.86}{3.86} & \stackcell{2.45}{3.45} &   &   &   & \stackcell{4.97}{5.96} & \stackcell{3.93}{4.89} & \stackcell{3.47}{4.44} & \stackcell{3.14}{4.09} & \stackcell{2.76}{3.71} \\[7pt]  
	  7 &   &   &   &   & \stackcell{6.05}{   7.00} & \stackcell{4.33}{ 5.40} & \stackcell{3.82}{4.88} & \stackcell{3.39}{4.44} &   &   &   &   & \stackcell{5.68}{6.74} & \stackcell{4.52}{5.53} & \stackcell{4.09}{5.11} & \stackcell{3.72}{4.78} \\[7pt]  
	  8 &   &   &   &   &   & \stackcell{6.78}{7.81} & \stackcell{4.93}{6.08} & \stackcell{4.35}{5.46} &   &   &   &   &   & \stackcell{6.42}{7.53} & \stackcell{5.14}{6.19} & \stackcell{4.68}{5.78} \\[7pt]  
	  9 &   &   &   &   &   &   & \stackcell{7.51}{8.63} & \stackcell{5.51}{6.74} &   &   &   &   &   &   & \stackcell{7.13}{8.27} & \stackcell{5.77}{6.91} \\[7pt]  
	  10 &   &   &   &   &   &   &   & \stackcell{8.21}{9.33} &   &   &   &   &   &   &   & \stackcell{7.84}{9.03} \\[7pt] 
	   \midrule
	\end{tabularx}
	\caption{Exact critical values of the test statistic $\text{LR}_{\bgamma^{(0)}}$ for static intensities subject to a significance level of $10\%$ (top number) and $5\%$ (bottom number)}
	\label{tab:crits-LR}
\end{table*}

\begin{table*}[tb]
\centering
\begin{tabularx}{\textwidth}{r|*{8}{X}|*{8}{X}}
 \toprule & \multicolumn{8}{c}{$\mathbf{M=5}$} &\multicolumn{8}{|c}{$\mathbf{M=10}$}\\ \midrule
$\mathbf{n}$ & 3 & 4 & 5 & 6 & 7 & 8 & 9 & 10 & 3 & 4 & 5 & 6 & 7 & 8 & 9 & 10 \\
  \midrule
  $\mathbf{r}$ & & & & & & & & & & & & & & & & \\
3 & \stackcell{9.61}{23.3} & \stackcell{10.6}{26.6} & \stackcell{11.5}{28.3} & \stackcell{11.6}{  29.0} & \stackcell{11.7}{  29.0} & \stackcell{11.7}{28.8} & \stackcell{12.3}{30.7} & \stackcell{12.3}{32.2} & \stackcell{2.26}{4.18} & \stackcell{2.14}{4.02} & \stackcell{2.15}{4.09} & \stackcell{2.17}{4.12} & \stackcell{2.18}{4.09} & \stackcell{2.18}{4.06} & \stackcell{2.16}{ 4.10} & \stackcell{2.15}{4.01} \\[7pt]  
  4 &   & \stackcell{14.7}{33.7} & \stackcell{15.2}{35.4} & \stackcell{16.2}{38.8} & \stackcell{16.6}{  40.0} & \stackcell{17.2}{41.7} & \stackcell{17.6}{43.9} & \stackcell{18.3}{46.2} &   & \stackcell{3.43}{6.17} & \stackcell{3.36}{6.06} & \stackcell{3.36}{6.09} & \stackcell{3.28}{5.95} & \stackcell{3.28}{5.93} & \stackcell{3.31}{5.93} & \stackcell{ 3.30}{5.91} \\[7pt]  
  5 &   &   & \stackcell{20.3}{44.4} & \stackcell{  21.0}{47.8} & \stackcell{21.9}{50.5} & \stackcell{22.3}{51.4} & \stackcell{22.7}{53.8} & \stackcell{23.2}{54.8} &   &   & \stackcell{4.72}{8.45} & \stackcell{4.58}{8.03} & \stackcell{4.51}{7.96} & \stackcell{4.57}{8.03} & \stackcell{ 4.5}{7.84} & \stackcell{4.46}{7.93} \\[7pt]  
  6 &   &   &   & \stackcell{25.7}{57.2} & \stackcell{26.6}{58.4} & \stackcell{27.8}{63.2} & \stackcell{28.5}{64.9} & \stackcell{29.1}{  68.0} &   &   &   & \stackcell{5.95}{10.7} & \stackcell{6.00}{10.1} & \stackcell{5.73}{10.1} & \stackcell{ 5.70}{ 9.90} & \stackcell{5.69}{9.88} \\[7pt]  
  7 &   &   &   &   & \stackcell{30.9}{68.8} & \stackcell{32.6}{72.8} & \stackcell{34.4}{75.8} & \stackcell{34.1}{77.1} &   &   &   &   & \stackcell{7.24}{12.8} & \stackcell{7.04}{12.4} & \stackcell{6.93}{12.1} & \stackcell{6.96}{12.1} \\[7pt]  
  8 &   &   &   &   &   & \stackcell{36.6}{80.5} & \stackcell{38.7}{85.2} & \stackcell{39.3}{87.6} &   &   &   &   &   & \stackcell{8.61}{15.2} & \stackcell{8.26}{14.6} & \stackcell{8.24}{14.3} \\[7pt]  
  9 &   &   &   &   &   &   & \stackcell{41.5}{92.7} & \stackcell{44.1}{97.2} &   &   &   &   &   &   & \stackcell{ 9.7}{17.2} & \stackcell{9.48}{16.9} \\[7pt]  
  10 &   &   &   &   &   &   &   & \stackcell{47.3}{ 103} &   &   &   &   &   &   &   & \stackcell{11.1}{19.7} \\[7pt] 
   \bottomrule
\end{tabularx}
\caption{Exact critical values of the test statistic $W_{\bgamma^{(0)}}$ for static intensities subject to a significance level of $10\%$ (top number) and $5\%$ (bottom number)}
\label{tab:crits-WP}
\end{table*}

\subsection{Testing the baseline distribution}\label{ss:gof}

As a step function,  the estimated cdf $\hat{F}_{\bgamma}$ is uniquely determined by its jump locations $X_i^{(j)}$ and the corresponding jump factors $Y_i^{(j)}$, $1\leq j\leq r$, $1\leq i\leq M$. By the same arguments as in Section \ref{ss:mc}, the values $Y_i^{(j)}$ depend on the observed failure times in $\mathbf{X}$ only via their ranks. Moreover, from the equivalence in (\ref{eq:F}) and equation (\ref{eq:Y}), we have for any (absolutely) continuous cdf $F$ that a.s.
\begin{equation*}
N_k(X_i^{(j)}-)\,=\,\sum_{l=1}^r \mathbf{1}_{\{U_k^{(l)} < U_i^{(j)}\}}\,,\quad 1\leq k\leq M\,,
\end{equation*}
and a.s.\
\begin{equation*}
X_i^{(j)}\leq t\quad\Leftrightarrow\quad U_i^{(j)}\leq F(t)
\end{equation*}
for $t>0$ and $1\leq j\leq r$, $1\leq i\leq M$. Hence, by using formula (\ref{eqn:F-kvam}),
\begin{equation}\label{eq:uni}
	\hat{F}_{\bgamma} (t)\,=\,1-\prod_{U_i^{(j)}\leq F(t)}(1-Y_i^{(j)}) \,=\, \hat{U}_{\bgamma}(F(t))\quad\text{a.s.}
\end{equation}
for $t>0$, where $\hat{U}_{\bgamma}$ denotes the Nelson-Aalen estimator based on $\mathbf{U}$. Since $\bU_1,\dots,\bU_M$ are distributed as $M$ iid vectors of the first $r\,(\leq n)$ SOSs based on the standard uniform cdf and parameters $\alpha_1,\dots,\alpha_n$, the distribution of $\hat{U}_{\bgamma}$ does not depend on $F$. This property remains true if $\bgamma$ is replaced by $\hat{\bgamma}^\text{PL}$ or $\hat{\bgamma}^\text{ML}$. 

Property \eqref{eq:uni} is analogous to the classical result for the empirical distribution function in the iid setting; see, e.g., \cite[Chapter 7]{shorack2000probability}. The benefit of relation (\ref{eq:uni}) is that it yields useful properties of the distribution of $\hat{F}_{\bgamma}$ and, in particular, allows for the construction of simple distribution-free test statistics, even for finite sample sizes. In the following, let $F^{-1}(0+)=\lim_{u\downarrow0}F^{-1}(u)$.

\begin{theorem}\label{thm:free2}
	For $q\in(0,1]$ and any continuous mapping $T:[0,1]\times (0,1)\to \R$, the statistic
\begin{align*}
		\sup_{F^{-1}(0+)< t<F^{-1}(q)} T(\hat{F}_{\bgamma}(t), F(t)) = \sup_{0< p< q} T(\hat{U}_{\bgamma}(p), p)
	\end{align*}
	is distribution-free w.r.t. $F\in\mathcal{F}$. Moreover, this property remains true if $\bgamma$ is replaced by $\hat{\bgamma}^{\text{PL}}$ or $\hat{\bgamma}^{\text{ML}}$.
\end{theorem}

In particular, Theorem \ref{thm:free2} yields that for $F\in\mathcal{F}$ with support $[0,\infty)$ the distribution of
\begin{align*}
		\sup_{0<t<\infty} T(\hat{F}_{\bgamma}(t), F(t)) \,=\, \sup_{0< p<1} T(\hat{U}_{\bgamma}(p), p)
	\end{align*}
does not depend on $F$. As a first example, this finding is applicable to the Kolmogorov-Smirnov type statistic defined by
\begin{equation*}
K_{\bgamma}\,=\,\sup_{0< t<\infty} |\hat{F}_{\bgamma}(t)-F(t)|
\end{equation*}
As a competing test statistic, we also address the weighted version 
\begin{align*}
\tilde{K}_{\bgamma}
	\,=\,\sup_{0<t<\infty} \frac{|\hat{F}_{\bgamma}(t) - F(t)|}{k_{\bgamma}(F(t))}
\end{align*}
with
\begin{align*}
	 k_{\bgamma}(p)\,&=\,(1-p)\sqrt{g_{\bgamma}(p)(1+|\ln (g_{\bgamma}(p))|)}\,,\quad p\in(0,1)\,,
\end{align*}
where $g_{\bgamma}(p) =
\int_0^p \lambda_U(s) / E(\gamma_{N(s)+1})\,  ds$, $p\in(0,1)$, denotes the asymptotic variance function of $\hat{U}_{\bgamma}(p)$ when $\bgamma$ is assumed to be known (see Theorem 3 in \cite{beutner2008nonparametric}). Here, the weighting scheme $k_{\bgamma}$ is motivated by the law of the iterated logarithm for the asymptotic Brownian motion, which ensures that $B(x)/\sqrt{x|\ln(x)|}\to 0$ for $x\to\infty$ and $x\downarrow0$. If $\gamma_1,\dots,\gamma_r$ are pairwise distinct, we have the explicit formula
\begin{align}
g_{\bgamma}(p)\,&=\,\int_0^p \Bigg(\sum_{j=1}^r (1-s)^{\gamma_j+1} \sum_{k=j}^r b_{j,k}(\bgamma) \Bigg)^{-1}\,ds,\label{eq:g}\\
\intertext{for $p\in(0,1)$, where for $1\leq j\leq k \leq r$,}
b_{j,k}(\bgamma)\,&=\,\Bigg(\prod_{l=1}^k \gamma_l \Bigg)\Bigg(\prod_{i=1, i\neq j}^k (\gamma_i-\gamma_j)\Bigg)^{-1}.\label{eq:b}
\end{align}
The derivation of this identity can be found in the appendix.

Another test statistic is proposed in \cite{beutner2008nonparametric}. Corresponding to $\hat{F}_{\bgamma}$ and with $Y_i^{(j)}$, $1\leq i\leq M$, $1\leq j\leq r$, as in formula (\ref{eqn:q}), the estimator \begin{align*}
	\hat{\Lambda}_{\bgamma}^F(t)=\sum_{X_i^{(j)}\leq t} Y_i^{(j)}\,,\quad t\geq0\,,
\end{align*}
for the cumulative hazard rate $\Lambda^F=-\ln(1-F)$ is employed to construct the test statistic process
\begin{align*}
	z_{\bgamma}^\rho(t)\,=\, \int_0^t (1-F(s))^\rho \sum_{i=1}^M \gamma_{N_i(s-)+1} \,d(\hat{\Lambda}_{\bgamma}^F-\Lambda^F)(s),
\end{align*}
for $t\geq 0$ and some $\rho\in[0,1]$. Using the same argument as in formula \eqref{eq:uni}, we obtain the representation 
\begin{align*}
	z_{\bgamma}^\rho(t)\,&=\, \int_0^{F(t)} (1-s)^{\rho}\sum_{i=1}^M \gamma_{N^U_i(s-)+1} \,d(\hat{\Lambda}_{\bgamma}^U-\Lambda^U)(s)\\
	&=\,\tilde{z}_{\bgamma}^\rho(F(t))\,,\quad t\geq0.
\end{align*}
Here, $\Lambda^U(s)=-\ln(1-s)$, $s\in[0,1)$, is the cumulative hazard rate of a standard uniform distribution, $N^U_i$ is the counting process with jump times $U_i^{(j)}$, $1\leq j\leq r$, in sample $i\in\{1,\dots,M\}$, $\hat{\Lambda}_{\bgamma}^U$ is the estimator based on $\bU_1,\dots,\bU_M$, and $\tilde{z}^\rho_{\bgamma}$ denotes the corresponding test statistic process. Thus, for $q\in(0,1]$, the statistic
\begin{align*}
\sup_{F^{-1}(0+)<t< F^{-1}(q)} |z^\rho_{\bgamma}(t)|\,=\,
\sup_{0<p<q} |\tilde{z}^\rho_{\bgamma}(p)|
\end{align*} 
is distribution-free w.r.t. $F\in\mathcal{F}$. In particular, we have
\begin{align*}
Z^\rho_{\bgamma}\,=\,\sup_{0<t<\infty} |z^\rho_{\bgamma}(t)|\,=\,\,\sup_{0<p<1} |\tilde{z}^\rho_{\bgamma}(p)|
\end{align*} 
for $F\in\mathcal{F}$ with support $[0,\infty)$.

If $\bgamma$ is known, the finite-sample distribution  of $K_{\bgamma}$, $\tilde{K}_{\bgamma}$, and $Z^\rho_{\bgamma}$ may be obtained computationally by means of Monte Carlo simulations. The resulting quantiles may then serve as critical values of the test statistics. Note that these values do not depend on the cdf specified under the null hypothesis.

\begin{remark}
As another application in reliability, our findings may also be useful in the context of progressively type-II censored lifetime experiments (see \cite{BalAgg2000,BalCra2014}). In such an experiment, $N$ components are put on a lifetime test. For $1\leq k\leq n-1$, upon the $k$-th component failure, $R_k\in\mathbb{N}_0$ of the still operating components are then randomly selected and removed from the experiment. After the $n$-th component failure, the remaining $R_n=N-n-\sum_{k=1}^{n-1}R_k$ components are removed. Hence, $n$ component failure times are recorded and $\sum_{k=1}^n R_k=N-n$ component failure times are progressively type-II censored.

Progressively type-II censored order statistics serve as a model for such lifetime experiments and are based on some underlying component lifetime cdf $F$, say, and the censoring scheme $(R_1,\dots,R_n)$. 
In distribution, progressively type-II censored order statistics based on some absolutely continuous cdf $F$ and censoring scheme $(R_1,\dots,R_n)$ coincide with SOSs based on $F$ and parameters
\begin{equation*}
\gamma_j=n-j+1+\sum_{k=j}^n R_k\,,\quad 1\leq j\leq n
\end{equation*}
(see, e.g., \cite[Section 2.2]{CraKam2001b} and \cite[Section 2.2]{BalCra2014}). In particular, the parameter vector $\bgamma$ is known, such that the preceding results provide exact goodness-of-fit tests for the underlying lifetime distribution of progressively type-II censored order statistics. For the discussion of goodness-of-fit testing with progressively type-II censored order statistic including test statistics based on spacings, the empirical distribution function, and divergence measures, we refer to \cite[Chapter 19]{BalCra2014} and the references therein.
\end{remark}

If $\bgamma$ is unknown, we might plug-in the profile likelihood estimator $\hat{\bgamma}^{\text{PL}}$, but the exact distributions of the above test statistics would still depend on $\bgamma$. In contrast, the MLE $\hat{\bgamma}^\text{ML}$ of $\bgamma$, which can be obtained from equation \eqref{eqn:alphamle}, is a sufficient statistic for $\bgamma$ if $F$ is fixed (see, e.g., \cite{cramer1996sequential}). The nuisance parameter $\bgamma$ can thus be handled by a test with Neyman structure as follows. By the quantile representation \eqref{eq:Y} and the previous discussion, the distribution of $(K_{\hat{\bgamma}^{\text{ML}}}, \hat{\bgamma}^{\text{ML}})$, and hence of $(K_{\hat{\bgamma}^{\text{ML}}} | \hat{\bgamma}^{\text{ML}})$, does not depend on $F$, where the latter denotes the conditional distribution of $K_{\hat{\bgamma}^{\text{ML}}}$ given $\hat{\bgamma}^{\text{ML}}$. By virtue of the sufficiency of $\hat{\bgamma}^{\text{ML}}$, this conditional distribution does not depend on $\bgamma$ either, such that conditional critical values $c(\hat{\bgamma}^{\text{ML}})$ for the test statistic $K_{\hat{\bgamma}^{\text{ML}}}$ can be determined. The same is true for $\tilde{K}_{\hat{\bgamma}^{\text{ML}}}$ and $Z^\rho_{\hat{\bgamma}^{\text{ML}}}$. In practice, the quantiles of the conditional distribution can be approximated to arbitrary precision by means of a Monte Carlo simulation. This yields an exact method for testing the hypothesis that $F$ is the true baseline cdf.

Since the conditional distribution of $K_{\hat{\bgamma}^{\text{ML}}}$ given $\hat{\bgamma}^{\text{ML}}$ does not depend on $F$, we may simplify computations of the conditional critical values by choosing a standard exponential baseline distribution. In the latter setting, the MLE has components
\begin{align*}
\hat{\gamma}_j^\text{ML}\,=\, M\left(\sum_{i=1}^M (X_i^{(j)}-X_i^{(j-1)})\right)^{-1}\,, \quad 2\leq j\leq n\,,
\end{align*}
and $\hat{\gamma}^\text{ML}_1=n$, since $\alpha_1=1$ by assumption (see formula (\ref{eqn:alphamle})).

\begin{theorem}\label{thm:conditional}
	Let $F$ be the standard exponential cdf and $\mathbf{X}=(X_i^{(j)})_{1\leq i\leq M,1\leq j\leq r}\in(0,\infty)^{M\times r}$ be the data matrix. Then, we have
	\begin{eqnarray*}
	\mathbf{X}\,|\,\hat{\bgamma}^\text{ML}\,&\thicksim&\,(\bV_1,\dots,\bV_r)\,\mathbf{diag}\left(\frac{1}{n},\frac{M}{\hat{\gamma}_2^{\text{ML}}},\dots,\frac{M}{\hat{\gamma}_r^{\text{ML}}}\right)\,\mathbf{A}\,,
\end{eqnarray*}
where $\mathbf{A}=(a_{ij})_{1\leq i,j\leq r}\in\R^{r\times r}$ is a deterministic matrix with entries $a_{ij}=1$ for $i\leq j$ and zero otherwise, and $\bV_1,\dots,\bV_r$ are independent $M$-dimensional random (column) vectors with
\begin{eqnarray*}
\bV_1\thicksim\bigotimes_{i=1}^M\exp(1)\quad\text{and}\quad \bV_j\thicksim\text{Dir}(1,\dots,1)\,,\quad 2\leq j\leq r\,.
\end{eqnarray*}
Here, $\exp(1)$ denote the standard exponential distribution and $\text{Dir}(1,\dots,1)$ a Dirichlet distribution whose $M$ parameters are all equal to 1.
\end{theorem}

\begin{figure*}[ht!]
\centering
	\includegraphics[width=0.45\textwidth]{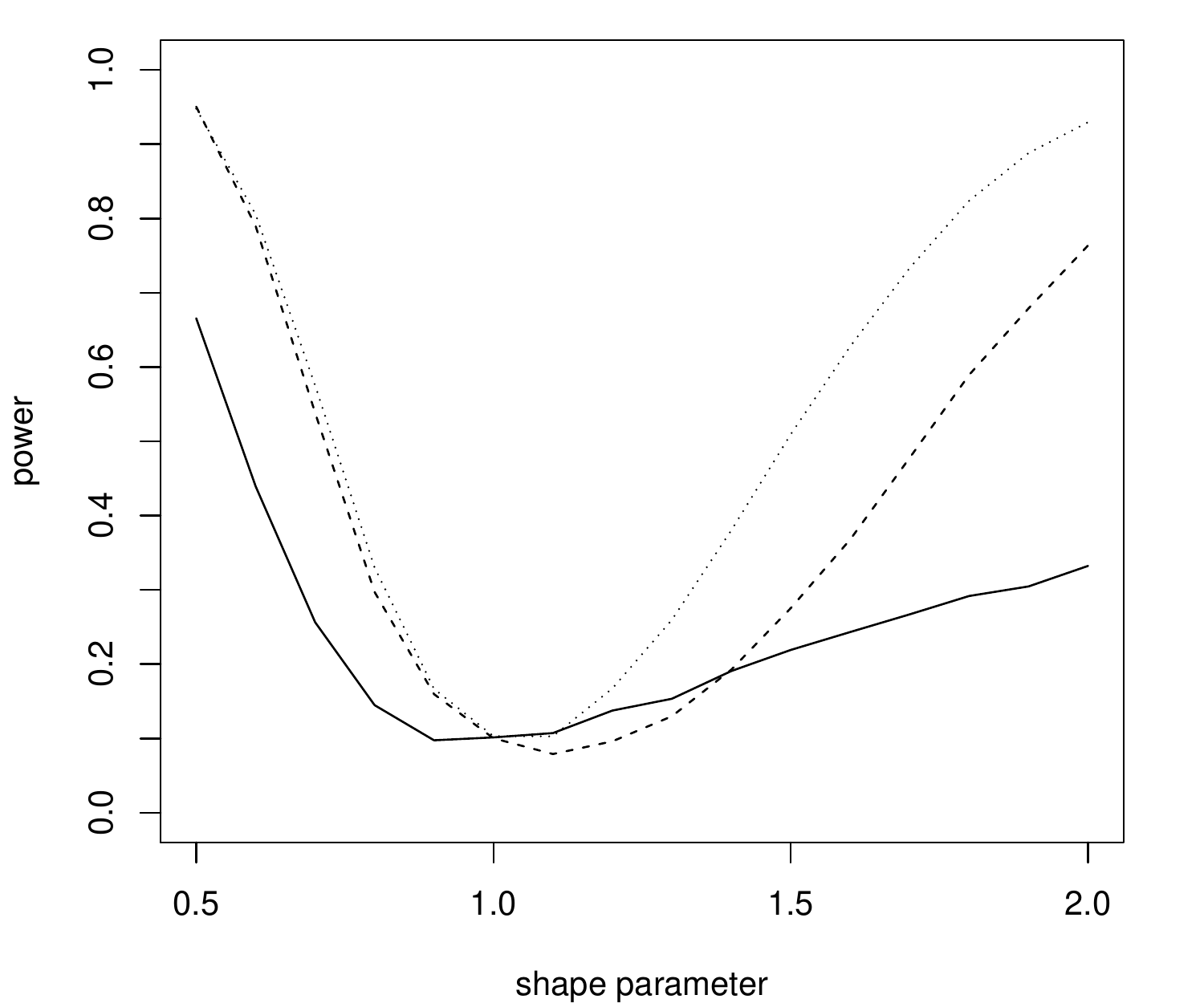}
	\includegraphics[width=0.45\textwidth]{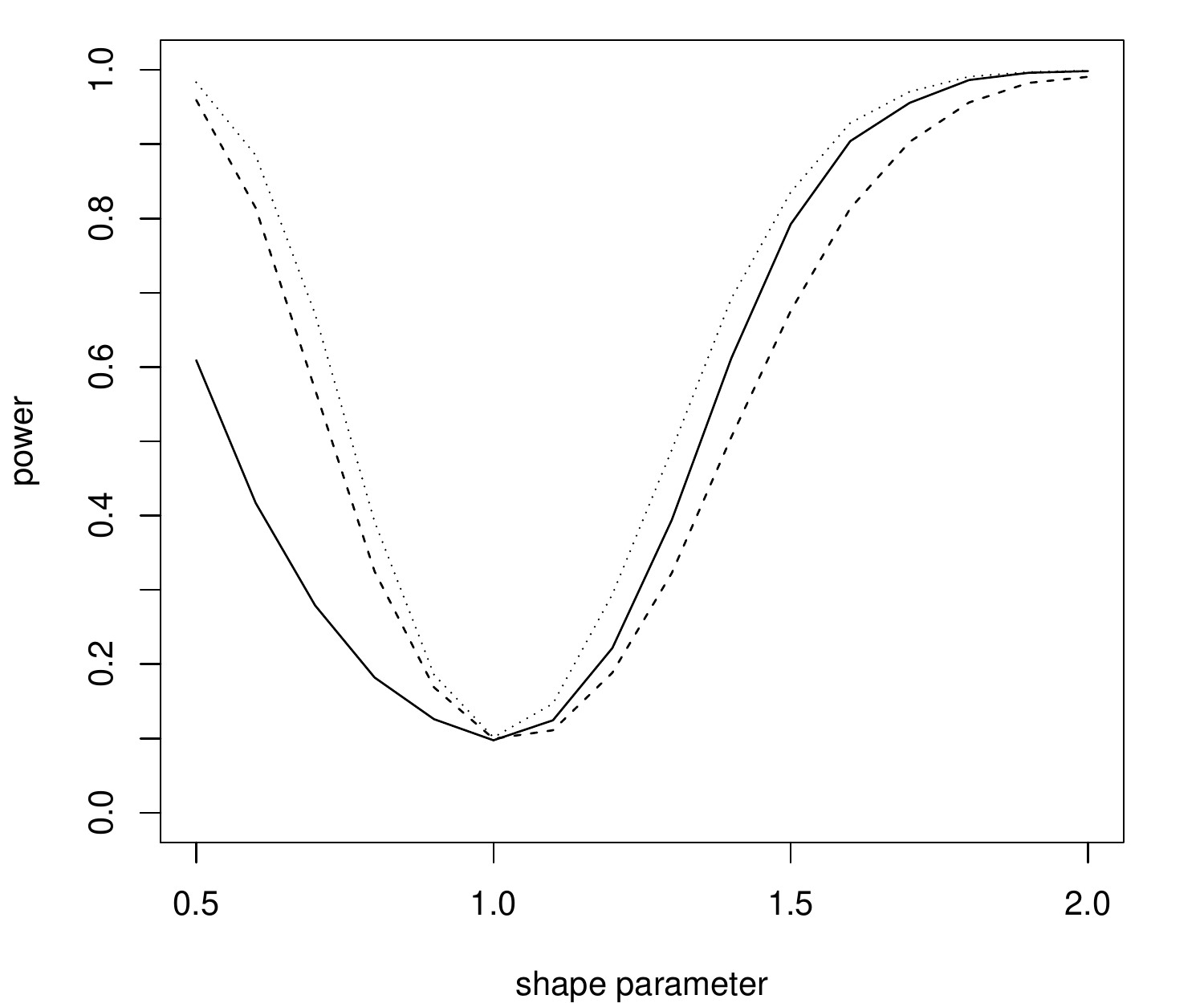}
	\caption{Power at Weibull distributions (left) and gamma distributions (right) with scale parameter 1 of the conditional tests based on $K_{\hat{\bgamma}^{\text{ML}}}$ (solid line), $\tilde{K}_{\hat{\bgamma}^{\text{ML}}}$ (dashed line), and $Z^{1/2}_{\hat{\bgamma}^{\text{ML}}}$ (dotted line) given $\hat{\bgamma}^{\text{ML}}$ when testing for a standard exponential baseline distribution at significance level 10\%, where $r=3$, $n=4$, $M=10$, and $\balpha = (1, 1.4, 1.8)$ respectively\ $\bgamma=(4,4.2,3.6)$.}
\label{fig:power-F}
\end{figure*}

The proof of Theorem \ref{thm:conditional} is presented in the appendix. It allows us to efficiently sample from the conditional distribution of $\mathbf{X}$ given $\hat{\bgamma}^\text{ML}$, and thus assess the conditional distributions of $K_{\hat{\bgamma}^{\text{ML}}}$, $\tilde{K}_{\hat{\bgamma}^{\text{ML}}}$, and $Z_{\hat{\bgamma}^{\text{ML}}}^\rho$ by means of Monte Carlo simulations. A respective quantile may then be used as a critical value in a conditional test of the hypothesis that $F$ is the true baseline cdf. For the null hypothesis of a standard exponential baseline distribution, a significance level of 10\%, and parameters $r=3$, $n=4$, $M=10$, and $\balpha=(1,1.4,1.8)$ respectively $\bgamma=(4,4.2,3.6)$, Figure \ref{fig:power-F} shows the power of the conditional tests based on $K_{\hat{\bgamma}^{\text{ML}}}$, $\tilde{K}_{\hat{\bgamma}^{\text{ML}}}$, and $Z_{\hat{\bgamma}^{\text{ML}}}^{1/2}$ at Weibull and gamma distributions with scale parameter 1, each. The conditional test based on $Z^{1/2}_{\hat{\bgamma}^{\text{ML}}}$ is found to perform best across these alternatives, being followed by the weighted Kolmogorov-Smirnov test based on $\tilde{K}_{\hat{\bgamma}^{\text{ML}}}$. All depicted power functions are based on $10^4$ realizations of $\bX$ and the corresponding values of $\hat{\bgamma}^{\text{ML}}$, for every underlying baseline distribution. Here, for every realization of $\hat{\bgamma}^{\text{ML}}$, the conditional critical value is, in turn, obtained from $10^2$ realizations of every test statistic under the null hypothesis, where we use the conditional distribution of $\mathbf{X}$ given $\hat{\bgamma}^{\text{ML}}$ as stated in Theorem \ref{thm:conditional}.

%--------------------

\section{Comparisons to asymptotic tests}\label{s:asym}

We compare the proposed exact statistical tests to their asymptotic variants based on the results in \cite{kvam2005estimating} and \cite{beutner2008nonparametric}. When testing for static intensities, the multivariate central limit theorem for $\hat{\bgamma}^{\text{PL}}$ can be used to derive asymptotic critical values for the Wald statistic $\text{W}_{\bgamma^{(0)}}$, with $\bgamma^{(0)}=(n,\ldots, n-r+1)$. The exact and the asymptotic procedure employ the same test statistic, and we investigate the size distortion incurred by using the asymptotic critical values. The corresponding simulation results are presented in Table \ref{tab:limit-alpha}. Here, as the distribution of $\hat{\bgamma}^{\text{PL}}$ does not depend on $F$, we use a standard uniform baseline cdf and set the truncation parameter in \cite{kvam2005estimating} to $\tau=0.98$. Based on the simulation results, it is found that the asymptotic test has a  relevant size distortion for small sample sizes, which suggests the use of exact critical values in practice.

\begin{table}[h!]
\centering
\begin{tabular}{r|rrrrrrr}
  \toprule
M & 5 & 10 & 15 & 20 & 30 & 40 & 50 \\ 
  \midrule
	size & .17 & .10 & .08 & .07 & .07 & .06 & .05\\
   \bottomrule
\end{tabular}
\caption{Size of the asymptotic Wald test based on $\text{W}_{\bgamma^{(0)}}$ with $\bgamma^{(0)}=(4,3,2,1)$ and a nominal level of $5\%$, where $n=r=4$ (obtained by $10^4$ Monte Carlo samples).}
\label{tab:limit-alpha}
\end{table}

To evaluate the tests for $F$, we choose a standard exponential baseline cdf and test against Weibull alternatives with scale parameter $1$ and different shape parameters. The asymptotic distribution of $K_{\hat{\bgamma}^{\text{PL}}}$,  $\tilde{K}_{\hat{\bgamma}^{\text{PL}}}$, and $Z^{1/2}_{\hat{\bgamma}^{\text{PL}}}$ is known from the results in \cite{kvam2005estimating} and \cite{beutner2008nonparametric}. It contains two orthogonal terms corresponding to the statistical error in the estimation of $F$ and $\bgamma$, respectively. The asymptotic variances depend on $\bgamma$, and we plug in $\hat{\bgamma}^{\text{PL}}$ to determine the critical values. In particular, these asymptotic tests are unconditional. The simulation results are presented in Table \ref{tab:limit-baseline} and indicate that this procedure leads to some size distortion. More prominently, the power of these asymptotic tests is not competitive, even for a large sample size of $M=50$. In further simulations not presented here, we correct for the size distortion of the asymptotic tests for unknown $\balpha$ by using the infeasible exact critical values. The latter simulations suggest that the power of the size-corrected tests is ultimately increasing in $M$, but still much lower than for the proposed conditional tests. This lack of power could be attributed to the error in estimating the nuisance parameter $\bgamma$, which leads to an additional asymptotic variance term dominating the test statistic. In fact, if we include $\bgamma$ as known and use the corresponding limit distributions, the size distortion of the asymptotic tests is negligible, and the power is closer to that of the proposed conditional tests (see Table \ref{tab:limit-baseline}). In practice, these asymptotic tests are infeasible if $\bgamma$ is unknown. Thus, the exact conditional tests compare favorably to the asymptotic benchmarks.

\begin{table*}[tb]
\centering
\begin{tabularx}{\textwidth}{cr|*{3}{X}|*{3}{X}|*{3}{X}}
  \toprule
  && \multicolumn{6}{c|}{$\balpha$ unknown} &  \multicolumn{3}{c}{$\balpha$ known} \\ \midrule
  shape&& \multicolumn{3}{c|}{exact tests based on} & \multicolumn{3}{c|}{asymptotic tests based on} & \multicolumn{3}{c}{asymptotic tests based on} \\
parameter & M & $K_{\hat{\bgamma}^{\text{ML}}}$ & $\tilde{K}_{\hat{\bgamma}^{\text{ML}}}$ & $Z^{1/2}_{\hat{\bgamma}^{\text{ML}}}$ & $K_{\hat{\bgamma}^{\text{PL}}}$ & $\tilde{K}_{\hat{\bgamma}^{\text{PL}}}$ & $Z^{1/2}_{\hat{\bgamma}^{\text{PL}}}$ & $K_{\bgamma}$ & $\tilde{K}_{\bgamma}$ & $Z^{1/2}_{\bgamma}$ \\ 
  \midrule
1.5 &  5 & .20 & .06 & .16 & .09 (.12) & .08 (.10) & .12 (.19) & .08 (.03) & .00 (.03) & .03 (.05) \\ 
    & 10 & .37 & .14 & .37 & .05 (.07) & .01 (.01) & .05 (.10) & .20 (.03) & .06 (.04) & .09 (.05) \\ 
    & 20 & .67 & .29 & .74 & .07 (.05) & .01 (.02) & .03 (.07) & .54 (.04) & .41 (.06) & .30 (.05) \\ 
    & 50 & .96 & .83 & .99 & .28 (.04) & .23 (.04) & .06 (.06) & .98 (.04) & .92 (.06) & .87 (.05) \\ \midrule
  0.8 & 5  & .07 & .16 & .15 & .15 (.12) & .12 (.10) & .23 (.19) & .06 (.03) & .12 (.03) & .12 (.05) \\ 
    & 10 & .12 & .20 & .23 & .09 (.07) & .04 (.01) & .13 (.10) & .10 (.03) & .15 (.04) & .15 (.05) \\ 
    & 20 & .28 & .27 & .39 & .08 (.05) & .05 (.02) & .11 (.07) & .17 (.04) & .21 (.06) & .21 (.05) \\ 
    & 50 & .64 & .52 & .72 & .09 (.04) & .08 (.04) & .10 (.06) & .46 (.04) & .36 (.06) & .39 (.05) \\ 
   \bottomrule
\end{tabularx}
\caption{Power of different tests for a standard exponential baseline cdf $F$ with a nominal level of $5\%$ against Weibull alternatives with selected shape parameters, where $\balpha = (1.0, 1.4, 1.8, 2.2)$ and $n=r=4$ (obtained by $10^4$ Monte Carlo samples). The size of the asymptotic tests is included in brackets.}
\label{tab:limit-baseline}
\end{table*}

%----------------------

\section{Real data example}\label{s:realdata}

We apply the tests of Section \ref{sec:model-check} to a ReliaSoft data set provided in \cite{Rel2002}, which was discussed in \cite{SutNai2014,KonYe2016}, and, more recently, in \cite{BedJohKam2019}. The data is shown in Table \ref{table:data_two_motors} and consists of the failure times $x_{i}^{(1)}<x_{i}^{(2)}$, $i=1,\dots,18$, of $n=2$ motors in $M=18$ identical parallel systems ($r=2$). In our model, the observations are assumed to be realizations of iid vectors $(X_i^{(1)},X_i^{(2)})$, $i=1,\dots,18$, of SOSs based on some absolutely continuous cdf $F$ and model parameter $\balpha=(1,\alpha_2)$ for some $\alpha_2>0$. %For a convenient scaling of the parameter estimates, we transform the data from days to years.

\begin{table}[h!]
\centering		
		\begin{tabular}{r|rrrrrrrrr}
			\toprule
				$i$ & 1 & 2 & 3 & 4 & 5 & 6  & 7 & 8 & 9\\ \midrule
				$x_i^{(1)}$ & 65 & 84 & 88 & 121 & 123 &  139 & 156 & 172 & 192\\[1ex]
				$x_i^{(2)}$ & 102 & 148 & 202 & 156 & 148 & 150 & 245 & 235 & 220 \\ \midrule\midrule
				$i$ & 10 & 11 & 12 & 13 & 14 & 15 & 16 & 17 & 18\\ \midrule
				$x_i^{(1)}$ & 207 & 212 &  212 & 213 & 220 & 243 & 248 & 257 & 263 \\[1ex]
				$x_i^{(2)}$ & 214 & 250 & 220 & 265 & 275 & 300 & 300 & 330 & 350\\
			\bottomrule
		\end{tabular}		
\caption{Failure times $x_{i}^{(1)}<x_{i}^{(2)}$ (in days) of $n=2$ motors in $M=18$ observed parallel systems (ReliaSoft 2002; see \cite{Rel2002}).}		\label{table:data_two_motors}
\end{table}
	
First, our aim is to decide, based on the given data, whether there is some load-sharing effect upon the first motor failure. For this, we compute the profile likelihood estimator $\hat{\balpha}^{\text{PL}}=(1,2.512)$ according to formula (\ref{eqn:gammaprofile}), indicating a load-sharing effect. To decide whether the load-sharing effect is significant, we apply the tests in Section \ref{ss:mc} to check for the null hypothesis
\begin{equation*}
H_0:\,\alpha_2=1\,.
\end{equation*}
The likelihood ratio test and the Wald test with test statistics as in formulas (\ref{eq:lrt}) and (\ref{eq:Wald}) with $\bgamma_0=(2,1)$ are found to have a $p$-value of 4\% and 1\%, respectively (obtained by $10^4$ Monte Carlo samples). Hence, based on the given data set, a load-sharing effect can be established subject to a significance level of 5\%; cf. \cite{SutNai2014,BedJohKam2019}.

\begin{figure}[ht!]
\centering
	\includegraphics[width=0.45\textwidth]{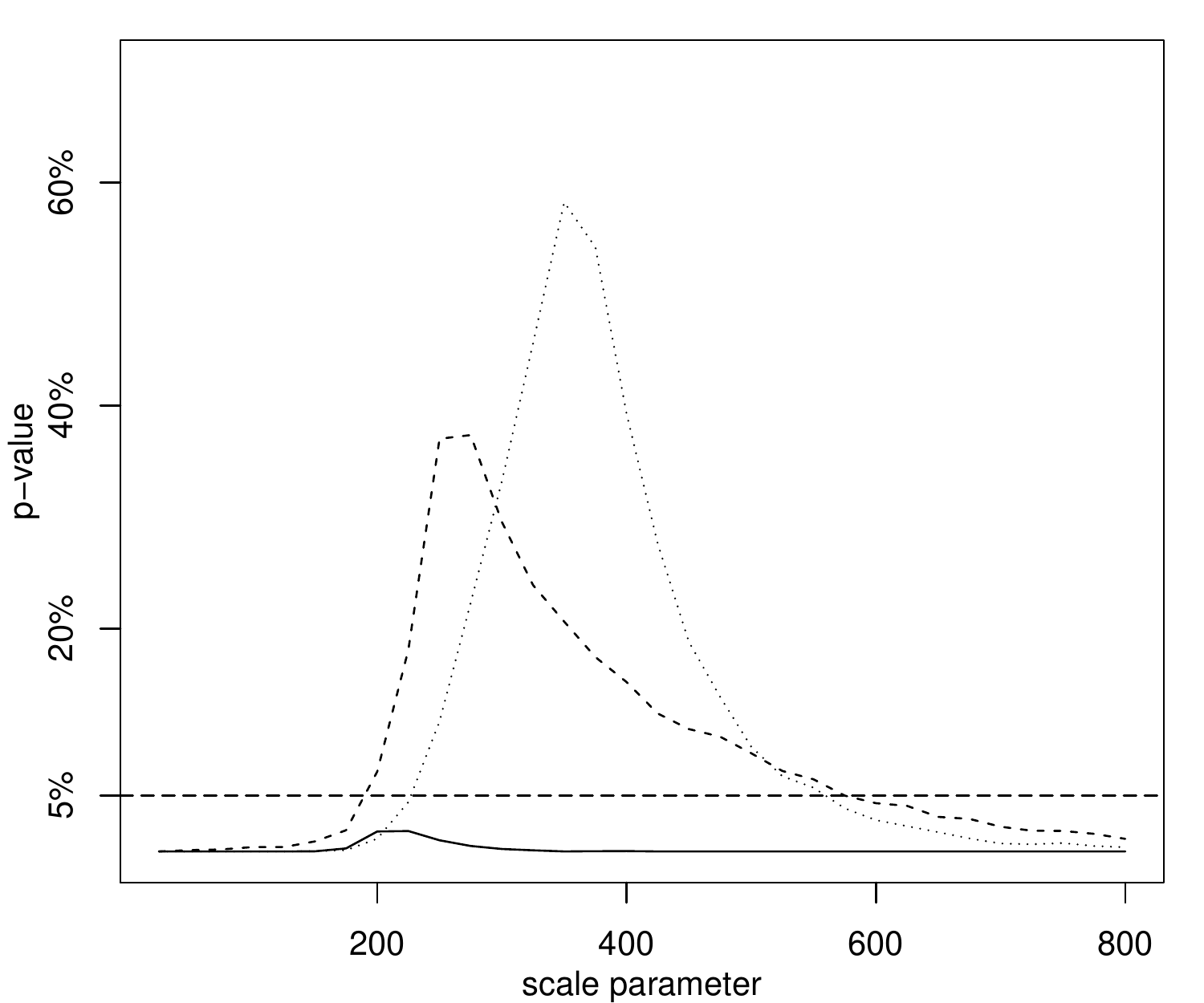}
\caption{$p$-values of the conditional tests based on $K_{\hat{\bgamma}^{\text{ML}}}$ (solid line), $\tilde{K}_{\hat{\bgamma}^{\text{ML}}}$ (dashed line), and $Z^{1/2}_{\hat{\bgamma}^{\text{ML}}}$ (dotted line) given $\hat{\bgamma}^{\text{ML}}$ when applied to the data in Table \ref{table:data_two_motors} and testing for null hypothesis (\ref{eq:realdataH02}) with $\sigma\in\{25,50,\dots,800\}$.}
\label{fig:real1}
\end{figure}

Next, we use the conditional tests proposed in Section \ref{ss:gof} to decide whether, based on the observed data, some particular exponential baseline distribution is adequate or has to be rejected for modeling. Here, we consider the null hypothesis
\begin{equation}\label{eq:realdataH02}
H_0:\,F=F_\sigma
\end{equation}
with exponential cdf
\begin{equation*}
F_\sigma(t)=1-\exp\left\{-\,\frac{t-50}{\sigma}\right\}\,,\quad t\geq50\,,
\end{equation*}
for some scale parameter $\sigma\in\{25,50,\dots,800\}$. Based on the data and $10^4$ Monte Carlo simulations from the conditional distribution of $\mathbf{X}$ given $\hat{\bgamma}^{\text{ML}}$ (for every value of $\sigma$), the $p$-values of the conditional tests are illustrated in Figure \ref{fig:real1} as a function of $\sigma$. Is is found that, for a significance level of 5\%, the conditional test based on $K_{\hat{\bgamma}^{\text{ML}}}$ rejects the null hypothesis (\ref{eq:realdataH02}) for all values of $\sigma$ considered, whereas the conditional tests based on $\tilde{K}_{\hat{\bgamma}^{\text{ML}}}$ and $Z^{1/2}_{\hat{\bgamma}^{\text{ML}}}$ do not reject $H_0$ for $\sigma\in\{200,225,\dots,575\}$ and $\sigma\in\{250,275,\dots,550\}$, respectively. In particular, based on the given observations, the conditional tests thus give evidence, subject to a significance level of 5\%, against the presence of an exponential distribution with location parameter $\mu=50$ and any fixed scale parameter $\sigma\in\{25,50,\dots,175\}$; cf. \cite{KonYe2016,BedJohKam2019}.

%-----------------

\section{Conclusion}\label{sec:concl}

We show that exact finite sample inference for sequential order statistics is feasible in a semiparametric framework, while maintaining statistical power. This comes at the cost that the relevant statistical distributions can no longer be handled analytically, but need to be assessed computationally. With modern computing power, however, this is not an issue in applications, although it might impede acceptance by practitioners. 

The presented approach could be transferred to related sequential order statistics models. In accelerated life-testing, for example, sequential order statistics serve as a step-stress model with proportional hazard rates and pre-specified numbers of observations under all stress levels (see \cite{BalKamKat2012,BedKamKat2015,BedKam2019}). Here, the model parameters associated with the stress levels are supposed to be connected via some parametric link function which allows for estimating the parameter corresponding to the stress under normal operating conditions based on data obtained from higher stress levels. Using the profile likelihood method, inference for the parameters of the link function should be unaffected by the nuisance baseline distribution.\\

\textbf{Acknowledgements}\\
The authors would like to thank the associate editor and three referees for their careful reading and constructive comments on the manuscript, which have led to improvements upon a previous version and to the inclusion of Sections \ref{s:asym} and \ref{s:realdata}.

\appendix

\subsection{Proof of formula (\ref{eq:g}).}

Let $\gamma_1,\dots,\gamma_r$ be pairwise distinct. We are working under the assumption of a standard uniform baseline, i.e.\ $\lambda_u(s)=1/(1-s)$ for $s\in(0,1)$. Moreover, for any $s\geq 0$, we have $N(s)=N(s-)$ almost surely, since the cumulative intensity process is absolutely continuous. By using formula (8) in \cite{KamCra2001} (see also \cite[Proposition 3.1]{jacobsen1984maximum}) then follows that, for $k\in\{0,1,\dots, r-1\}$,
\begin{align*}
	&\qquad P(N(s)=k)\\
	&=\,\left(\prod_{l=1}^k\gamma_l\right)\,\sum_{j=1}^{k+1} \left\{ (1-s)^{\gamma_j} \Bigg(\prod_{i=1, i\neq j}^{k+1} (\gamma_i-\gamma_j) \Bigg)^{-1}\right\}\,,
\end{align*}
and, hence,
\begin{equation*}
	\gamma_{k} \,P(N(s)=k-1) \,=\, \sum_{j=1}^k (1-s)^{\gamma_j}\, b_{j,k}(\bgamma)\,,\quad 1\leq k\leq r\,,
\end{equation*}
with $b_{j,k}(\bgamma)$, $1\leq j\leq k\leq r$, as in formula (\ref{eq:b}). Thus, since $\gamma_{r+1}=0$,
\begin{eqnarray*}
	E(\gamma_{N(s)+1}) \,&=&\, \sum_{k=1}^{r} \gamma_k P(N(s)=k-1) \\
	&=&\, \sum_{j=1}^{r}  (1-s)^{\gamma_j} \sum_{k=j}^r b_{j,k}(\bgamma)\,,
\end{eqnarray*}
and formula (\ref{eq:g}) is established.

\subsection{Proof of Theorem \ref{thm:conditional}}

From Section \ref{sec:model-spec}, the random $(M\times r)$-matrix $\mathbf{S}=(\bS^{(1)},\dots,\bS^{(r)})$ with column vectors
$\bS^{(j)}=(S_1^{(j)},\dots,S_M^{(j)})$, $1\leq j\leq r$, consists of independent random variables, where $S_1^{(j)},\dots,S_M^{(j)}$ are exponentially distributed with mean $1/\alpha_j$ for $1\leq j\leq r$. Moreover, for $2\leq j\leq r$, we introduce the random column vector $\tilde{\bS}^{(j)}=(S_1^{(j)},\dots, S_{M-1}^{(j)})$ and the random variable $S_\bullet^{(j)}=\sum_{i=1}^MS_i^{(j)}$ which has a gamma distribution with shape parameter $M$ and scale parameter $1/\alpha_j$. 
Then, for $\tilde{\boldsymbol{s}}^{(j)}=(\tilde{s}_1^{(j)},\dots, \tilde{s}_{M-1}^{(j)})\in(0,\infty)^{M-1}$, $2\leq j\leq r$, and $s_\bullet^{(2)},\dots,s_\bullet^{(r)}>0$,
\begin{eqnarray*}
	&&\quad f^{(\tilde{\bS}^{(2)},\dots,\tilde{\bS}^{(r)})\,|\,S_\bullet^{(j)}=s_\bullet ^{(j)},\,2\leq j\leq r}(\tilde{\bs}^{(2)},\dots,\tilde{\bs}^{(r)})\\
	&=&\,\prod_{j=2}^r \frac{f^{(\tilde{\bS}^{(j)},S_\bullet^{(j)})}(\tilde{\bs}^{(j)},s_\bullet ^{(j)})}{f^{S_\bullet^{(j)}}(s_\bullet^{(j)})}\\[1ex]
	&=&\,\prod_{j=2}^r \frac{f^{\bS^{(j)}}(\tilde{\bs}^{(j)},s_\bullet ^{(j)}-\sum_{i=1}^{M-1}\tilde{s}_i^{(j)})}{f^{S_\bullet^{(j)}}(s_\bullet^{(j)})}\\[1ex]
	&=&\,\prod_{j=2}^r \frac{(M-1)!}{\big(s_\bullet^{(j)}\big)^{M-1}}\,\boldsymbol{1}_{(0,s_\bullet^{(j)})}\left(\sum_{i=1}^{M-1} \tilde{s}_i^{(j)}\right)\,,
\end{eqnarray*}
by using the density transformation theorem and then the representations of the densities of exponential and gamma distributions. Note that the term behind the last product sign is the density of an $m$-dimensional Dirichlet distribution scaled by $s_\bullet^{(j)}$. Hence, we have
\begin{align*}
	(\bS^{(2)},\dots,\bS^{(r)})\,|\,(S_\bullet^{(2)},\dots,S_\bullet^{(r)})\,\thicksim\, (S_\bullet^{(2)}\bV_2,\dots,S_\bullet^{(r)}\bV_r),
\end{align*}
and, thus,
\begin{align*}
	\mathbf{S}\,|\,(S_\bullet^{(2)},\dots,S_\bullet^{(r)})\,\thicksim\,(\bV_1,\dots,\bV_r)\,\mathbf{diag}(1,S_\bullet^{(2)},\dots,S_\bullet^{(r)})
\end{align*}
with the $\bV$'s as stated in the theorem. Now, we rewrite $S_\bullet^{(j)}=M(n-j+1)/\hat{\gamma}_j^{\text{ML}}$, $2\leq j\leq r$, and, since for an exponential baseline cdf $\mathbf{X}=\mathbf{S}\mathbf{B}$ with matrix $\mathbf{B}\in\R^{r\times r}$ having column vectors $b_j=(1/n,\dots,1/(n-j+1),0\dots,0)$, $1\leq j\leq r$, the proof is completed.

\bibliographystyle{IEEEtran}
\bibliography{SOS_modelchecking}

\begin{thebibliography}{10}
\providecommand{\url}[1]{#1}
\csname url@samestyle\endcsname
\providecommand{\newblock}{\relax}
\providecommand{\bibinfo}[2]{#2}
\providecommand{\BIBentrySTDinterwordspacing}{\spaceskip=0pt\relax}
\providecommand{\BIBentryALTinterwordstretchfactor}{4}
\providecommand{\BIBentryALTinterwordspacing}{\spaceskip=\fontdimen2\font plus
\BIBentryALTinterwordstretchfactor\fontdimen3\font minus
  \fontdimen4\font\relax}
\providecommand{\BIBforeignlanguage}[2]{{%
\expandafter\ifx\csname l@#1\endcsname\relax
\typeout{** WARNING: IEEEtran.bst: No hyphenation pattern has been}%
\typeout{** loaded for the language `#1'. Using the pattern for}%
\typeout{** the default language instead.}%
\else
\language=\csname l@#1\endcsname
\fi
#2}}
\providecommand{\BIBdecl}{\relax}
\BIBdecl

\bibitem{kamps1995concept}
U.~Kamps, ``A concept of generalized order statistics,'' \emph{Journal of
  Statistical Planning and Inference}, vol.~48, no.~1, pp. 1--23, 1995.

\bibitem{Kam1995b}
------, \emph{A Concept of Generalized Order Statistics}.\hskip 1em plus 0.5em
  minus 0.4em\relax Stuttgart: Teubner, 1995.

\bibitem{CraKam2001b}
E.~Cramer and U.~Kamps, ``Sequential $k$-out-of-$n$ systems,'' in
  \emph{Handbook of Statistics, Advances in Reliability}, N.~Balakrishnan and
  C.~R. Rao, Eds.\hskip 1em plus 0.5em minus 0.4em\relax Amsterdam: Elsevier,
  2001, vol.~20, pp. 301--372.

\bibitem{BalBeuKam2008}
N.~Balakrishnan, E.~Beutner, and U.~Kamps, ``Order restricted inference for
  sequential $k$-out-of-$n$ systems,'' \emph{Journal of Multivariate Analysis},
  vol.~99, no.~7, pp. 1489--1502, 2008.

\bibitem{BalBeuKam2011}
------, ``Modeling parameters of a load-sharing system through link functions
  in sequential order statistics models and associated inference,'' \emph{IEEE
  Transactions on Reliability}, vol.~60, no.~3, pp. 605--611, 2011.

\bibitem{BedBeuKam2012}
S.~Bedbur, E.~Beutner, and U.~Kamps, ``Generalized order statistics: an
  exponential family in model parameters,'' \emph{Statistics}, vol.~46, no.~2,
  pp. 159--166, 2012.

\bibitem{BedBurKam2016}
S.~Bedbur, M.~Burkschat, and U.~Kamps, ``Inference in a model of successive
  failures with shape-adjusted hazard rates,'' \emph{Annals of the Institute of
  Statistical Mathematics}, vol.~68, no.~3, pp. 639--657, 2016.

\bibitem{BedLenKam2013}
S.~Bedbur, J.~M. Lennartz, and U.~Kamps, ``Confidence regions in models of
  ordered data,'' \emph{Journal of Statistical Theory and Practice}, vol.~7,
  no.~1, pp. 59--72, 2013.

\bibitem{BeuKam2009}
E.~Beutner and U.~Kamps, ``Order restricted statistical inference for scale
  parameters based on sequential order statistics,'' \emph{Journal of
  Statistical Planning and Inference}, vol. 139, no.~9, pp. 2963--2969, 2009.

\bibitem{cramer1996sequential}
E.~Cramer and U.~Kamps, ``Sequential order statistics and $k$-out-of-$n$
  systems with sequentially adjusted failure rates,'' \emph{Annals of the
  Institute of Statistical Mathematics}, vol.~48, no.~3, pp. 535--549, 1996.

\bibitem{CraKam2001a}
------, ``Estimation with sequential order statistics from exponential
  distributions,'' \emph{Annals of the Institute of Statistical Mathematics},
  vol.~53, no.~2, pp. 307--324, 2001.

\bibitem{cox1972regression}
D.~R. Cox, ``Regression models and life-tables,'' \emph{Journal of the Royal
  Statistical Society. Series B (Methodological)}, vol.~34, no.~2, pp.
  187--220, 1972.

\bibitem{aalen1978nonparametric}
O.~Aalen, ``Nonparametric inference for a family of counting processes,''
  \emph{The Annals of Statistics}, vol.~6, no.~4, pp. 701--726, 1978.

\bibitem{jacobsen1984maximum}
M.~Jacobsen, ``Maximum likelihood estimation in the multiplicative intensity
  model: a survey,'' \emph{International Statistical Review}, vol.~52, no.~2,
  pp. 193--207, 1984.

\bibitem{kvam2005estimating}
P.~H. Kvam and E.~A. Pe$\tilde{\text{n}}$a, ``Estimating load-sharing
  properties in a dynamic reliability system,'' \emph{Journal of the American
  Statistical Association}, vol. 100, no. 469, pp. 262--272, 2005.

\bibitem{beutner2008nonparametric}
E.~Beutner, ``Nonparametric inference for sequential $k$-out-of-$n$ systems,''
  \emph{Annals of the Institute of Statistical Mathematics}, vol.~60, no.~3,
  pp. 605--626, 2008.

\bibitem{Beu2010b}
------, ``Nonparametric comparison of several $k$-out-of-$n$ systems,'' in
  \emph{Advances in Data Analysis, Statistics for Industry and Technology},
  C.~Skiadas, Ed.\hskip 1em plus 0.5em minus 0.4em\relax Boston: Birkh\"auser,
  2010, pp. 291--304.

\bibitem{beutner2010nonparametric}
------, ``Nonparametric model checking for $k$-out-of-$n$ systems,''
  \emph{Journal of Statistical Planning and Inference}, vol. 140, no.~3, pp.
  626--639, 2010.

\bibitem{CraKam2003}
E.~Cramer and U.~Kamps, ``Marginal distributions of sequential and generalized
  order statistics,'' \emph{Metrika}, vol.~58, no.~3, pp. 293--310, 2003.

\bibitem{jacod1975multivariate}
J.~Jacod, ``Multivariate point processes: predictable projection,
  {R}adon-{N}ikodym derivatives, representation of martingales,''
  \emph{Probability Theory and Related Fields}, vol.~31, no.~3, pp. 235--253,
  1975.

\bibitem{PenHol2004}
E.~Pe$\tilde{\text{n}}$a and M.~Hollander, ``Models for recurrent events in
  reliability and survival analysis,'' in \emph{Mathematical Reliability: An
  Expository Perspective}, R.~Soyer, T.~Mazzuchi, and N.~Singpurwalla,
  Eds.\hskip 1em plus 0.5em minus 0.4em\relax Springer, 2004, vol.~67, pp.
  105--123.

\bibitem{PenSlaGon2007}
E.~A. Pe$\tilde{\text{n}}$a, E.~H. Slate, and J.~R. Gonzalez, ``Semiparametric
  inference for a general class of models for recurrent events,'' \emph{Journal
  of Statistical Planning and Inference}, vol. 137, no.~6, pp. 1727--1747,
  2007.

\bibitem{Pen2016}
E.~A. Pe$\tilde{\text{n}}$a, ``Asymptotics for a class of dynamic recurrent
  event models,'' \emph{Journal of Nonparametric Statistics}, vol.~28, no.~4,
  pp. 716--735, 2016.

\bibitem{BeuBorDoy2017}
E.~Beutner, L.~Bordes, and L.~Doyen, ``The failure of the profile likelihood
  method for a large class of semi-parametric models,'' \emph{Bernoulli},
  vol.~23, no.~4B, pp. 3650--3684, 2017.

\bibitem{KamCra2001}
U.~Kamps and E.~Cramer, ``On distributions of generalized order statistics,''
  \emph{Statistics}, vol.~35, no.~3, pp. 269--280, 2001.

\bibitem{BedBeuKam2014}
S.~Bedbur, E.~Beutner, and U.~Kamps, ``Multivariate testing and model-checking
  for generalized order statistics with applications,'' \emph{Statistics},
  vol.~48, no.~6, pp. 1297--1310, 2014.

\bibitem{shorack2000probability}
G.~R. Shorack, \emph{Probability for Statisticians}.\hskip 1em plus 0.5em minus
  0.4em\relax Springer, 2000.

\bibitem{BalAgg2000}
N.~Balakrishnan and R.~Aggarwala, \emph{Progressive Censoring: Theory, Methods,
  and Applications}.\hskip 1em plus 0.5em minus 0.4em\relax Boston:
  Birkh\"auser, 2000.

\bibitem{BalCra2014}
N.~Balakrishnan and E.~Cramer, \emph{The Art of Progressive Censoring}.\hskip
  1em plus 0.5em minus 0.4em\relax New York: Birkh\"auser, 2014.

\bibitem{Rel2002}
{ReliaSoft R\&D staff}, ``Using {QALT} models to analyze system configurations
  with load sharing,'' \emph{ReliaSoft's Reliability Edge Newsletter}, vol.~3,
  no.~3, pp. 1--4, 2002.

\bibitem{SutNai2014}
S.~Sutar and U.~Naik-Nimbalkar, ``Accelerated failure time models for load
  sharing systems,'' \emph{IEEE Transactions on Reliability}, vol.~63, no.~3,
  pp. 706--714, 2014.

\bibitem{KonYe2016}
Y.~Kong and Z.~Ye, ``A cumulative-exposure-based algorithm for failure time
  data from a load-sharing system,'' \emph{IEEE Transactions on Reliability},
  vol.~65, no.~2, pp. 1001--1013, 2016.

\bibitem{BedJohKam2019}
S.~Bedbur, M.~Johnen, and U.~Kamps, ``Inference from multiple samples of
  {W}eibull sequential order statistics,'' \emph{Journal of Multivariate
  Analysis}, vol. 169, pp. 381--399, 2019.

\bibitem{BalKamKat2012}
N.~Balakrishnan, U.~Kamps, and M.~Kateri, ``A sequential order statistics
  approach to step-stress testing,'' \emph{Annals of the Institute of
  Statistical Mathematics}, vol.~64, no.~2, pp. 303--318, 2012.

\bibitem{BedKamKat2015}
S.~Bedbur, U.~Kamps, and M.~Kateri, ``Meta-analysis of general step-stress
  experiments under repeated type-{II} censoring,'' \emph{Applied Mathematical
  Modelling}, vol.~39, no.~8, pp. 2261--2275, 2015.

\bibitem{BedKam2019}
S.~Bedbur and U.~Kamps, ``Confidence regions in step-stress experiments with
  multiple samples under repeated type-{II} censoring,'' \emph{Statistics \&
  Probability Letters}, vol. 146, pp. 181--186, 2019.

\end{thebibliography}

\end{document}